# LOPES 3D – studies on the benefits of EAS-radio measurements with vertically aligned antennas

D. Huber[1*], W.D. Apel[*], J.C. Arteaga-Velazquez[†], L. Bähren[**], K. Bekk[*],
M. Bertaina[‡], P.L. Biermann[§], J. Blümer[*], H. Bozdog[*], I.M. Brancus[¶], E. Cantoni[‡],
A. Chiavassa[‡], K. Daumiller[*], V. de Souza[∥], F. Di Pierro[‡], P. Doll[*], R. Engel[*],
H. Falcke[**,††], B. Fuchs[*], D. Fuhrmann[‡‡], H. Gemmeke[*], C. Grupen[§§], A. Haungs[*],
D. Heck[*], J.R. Hörandel[**], A. Horneffer[§], T. Huege[*], P.G. Isar[¶¶], K.-H. Kampert[‡‡],
D. Kang[*], O. Krömer[*], J. Kuijpers[**], K. Link[*], P. Łuczak[***], M. Ludwig[*],
H.J. Mathes[*], M. Melissas[*], C. Morello[†††], J. Oehlschläger[*], N. Palmieri[*],
T. Pierog[*], J. Rautenberg[‡‡], H. Rebel[*], M. Roth[*], C. Rühle[*], A. Saftoiu[¶],
H. Schieler[*], A. Schmidt[*], S. Schoo[*], F.G. Schröder[*], O. Sima[‡‡‡], G. Toma[¶],
G.C. Trinchero[†††], A. Weindl[*], J. Wochele[*], J. Zabierowski[***] and J.A. Zensus[§]

[*]*Karlsruhe Institute of Technology (KIT), Germany*
[†]*Universidad Michoacana, Morelia, Mexico*
[**]*Radboud University Nijmegen, Department of Astrophysics, The Netherlands*
[‡]*Dipartimento di Fisica dell' Università Torino, Italy*
[§]*Max-Planck-Institut für Radioastronomie Bonn, Germany*
[¶]*National Institute of Physics and Nuclear Engineering, Bucharest, Romania*
[∥]*Universidad São Paulo, Inst. de Física de São Carlos, Brasil*
[††]*ASTRON, Dwingeloo, The Netherlands*
[‡‡]*Universität Wuppertal, Fachbereich Physik, Germany*
[§§]*Universität Siegen, Fachbereich Physik, Germany*
[¶¶]*Institute for Space Sciences, Bucharest, Romania*
[***]*National Centre for Nuclear Research, Department of Astrophysics, Łódź, Poland*
[†††]*INAF Torino, Osservatorio Astrofisico di Torino, Italy*
[‡‡‡]*University of Bucharest, Department of Physics, Romania*

**Abstract.** The LOPES experiment was a radio interferometer built at the existing air shower array KASCADE-Grande in Karlsruhe, Germany. The last configuration of LOPES was called LOPES 3D and consisted of ten tripole antennas. Each of these antennas consisted of three crossed dipoles east-west, north-south, and vertically aligned. With this, LOPES 3D had the unique possibility to study the benefits of measurements with vertically aligned antennas in the environment of the well understood and calibrated particle detector array KASCADE-Grande. The measurements with three spatially coincident antennas allows a redundant reconstruction of the electric field vector. Several methods to exploit the redundancy were developed and tested. Furthermore, for the first time in LOPES, the background noise could be studied polarization- and direction dependent. With LOPES 3D it could be demonstrated that radio detection reaches a higher efficiency for inclined showers when including measurements with vertically aligned antennas and that the vertical component gets more important for the measurement of inclined showers. In this contribution we discuss a weighting scheme for the best combination of three redundant reconstructed electric field vectors. Furthermore, we discuss the influence of these weighting schemes on the ability to reconstruct air showers using the radio method. We show an estimate of the radio efficiency for inclined showers with focus on the benefits of measurements with vertically aligned antennas and we present the direction dependent noise in the different polarizations.

[1] Corresponding author
Email address: daniel.huber@kit.edu

# INTRODUCTION

The radio detection [1, 2, 3] of air showers is so far the only method combining a high uptime with a precise measurement of $X_{max}$ [4, 5]. Thus the interest in further developing this technique and gaining the maximum precision is very high. With LOPES 3D [6] the benefits of additional direct measurements with vertically aligned antennas have been studied. In principle, with measurements of two antennas at the same location the complete electric field vector can be reconstructed when knowing the arrival direction of the shower. This is already done by radio experiments like AERA [7] and Tunka-Rex [8]. LOPES 3D measures air showers with three antennas at the same location, thus the same electric field vector can be reconstructed three times. Several methods to make use of this redundancy were developed and one of them is discussed here exemplarily. These methods can be adopted out of the box for other radio experiments that are measuring three polarisation directions. In addition, the direction dependent background noise was studied in the single components separately.

# LOPES 3D

With LOPES 3D we wanted to study the benefits of measurements with vertically aligned antennas. Therefore, the LOPES experiment [9] was reconfigured with tripole antennas [10] as mentioned above and described in [6]. We have reconfigured and calibrated the complete experiment and developed a new timing calibration method [11] as well as a method to perform an amplitude calibration on vertically aligned antennas. It was shown that LOPES 3D is sensitive to the transit of the galactic plane and that solar flares can be detected with LOPES 3D [11]. However, the analysis of air shower events was harmed by several aspects:

- Increase in the background noise: see figure 1, where the average background noise is shown as a function of time on the left hand side. On the right hand side the detected and reconstructed events from LOPES 3D and the received triggers for the same time period are shown. It can be seen that after the increase in background noise a fewer number of events was detected. In fact the rate is down by a factor of 6. Thus the analysis performed and discussed in the following is harmed in the sense that no statistically significant conclusion can be given.
- Reduction of antenna positions: We reduced the setup from 25 antenna positions to ten. Consequently the efficiency of the antenna array went down roughly by a factor of 2 [6]
- Incorporation of the vertical component, where in most cases only a weak signal but high noise is expected. Measuring also with vertically aligned antennas was the goal and motivation for LOPES 3D, but nevertheless this harms the data quality compared to the former setups of LOPES.

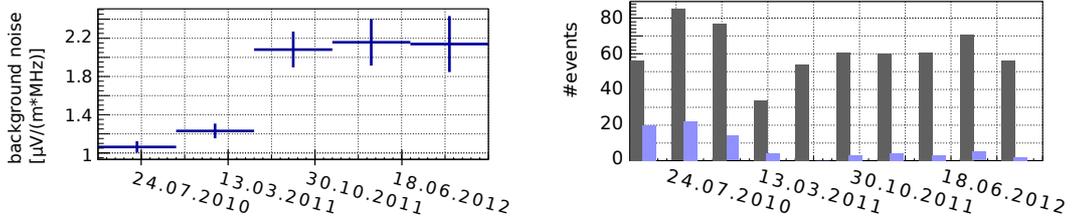

**FIGURE 1.** Average background noise as a function of time (lhs). Detected and reconstructed air showers from LOPES 3D in blue and received triggers in black as a function of time (rhs). The observed decrease in the event rate is caused by the increase of the background noise, while the trigger is constant.

The methods developed within LOPES 3D and the knowledge derived are partly applied at current radio observatories like AERA and Tunka-Rex. In the following we present the latest results retrieved by analysing the LOPES 3D data taken before the increase in background noise.

# INCORPORATION OF THE ANTENNA GAIN

During the measurement the electric field vector is converted to a voltage at the antenna foot-point. Thus, the vectorial information gets lost, as a vector is converted to a scalar, $S_{ant} = \vec{E} \cdot \vec{G}(f, \theta, \phi)$, with the electric field vector $\vec{E}$, the gain of the antenna $\vec{G}(f, \theta, \phi)$, the frequency of the signal $f$, the azimuth $\phi$, the zenith angle $\theta$, and the resulting voltage at the antenna foot-point $S_{ant}$. It is obvious that the electric field vector cannot be reconstructed when using measurements from only one antenna. In principle three independent measurements are needed in order to fully reconstruct the electric field vector as the gain and the voltages at the antenna foot-points are known. Thus the resulting linear equation system $\vec{E} \cdot \vec{G}_{antx} = S_{antx}$ ($x = 1,2,3$ for the 3 different measurements) can be solved for $\vec{E}$. Like any electromagnetic wave, the radio emission from cosmic ray induced air showers is a transverse wave, consequently the component of the electric field vector along the shower axis equals zero. In this frame a measurement with two antennas in spatial coincidence is sufficient to reconstruct the electric field vector according to equation 1 with the two components of the electric field vector in the new reference frame $E_\theta$ and $E_\phi$.

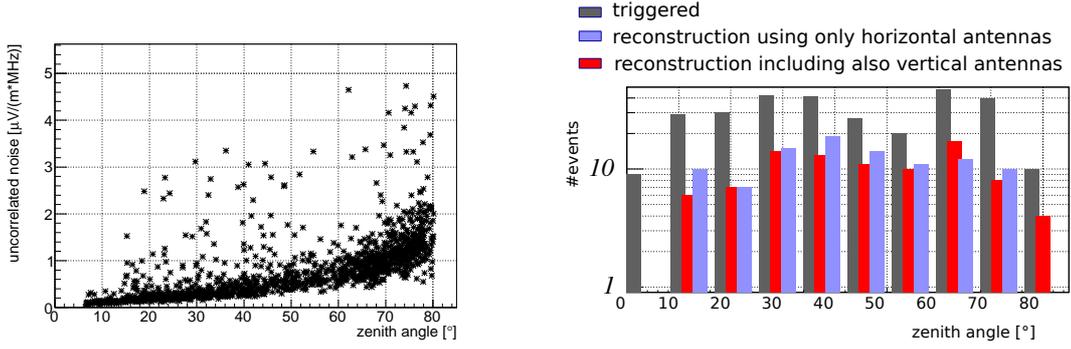

**FIGURE 2.** Background noise in vertical component as a function of zenith angle (lhs). Absolute amount of reconstructed events for KASCADE, black, for LOPES when only using horizontally aligned antennas, blue, and when including also vertically aligned antennas red (rhs).

$$E_\theta = \frac{S_{ant2} \cdot G^{az}_{ant1} - S_{ant1} \cdot G^{az}_{ant2}}{G^{az}_{ant1} \cdot G^{ze}_{ant2} - G^{az}_{ant2} \cdot G^{ze}_{ant1}}, \qquad E_\phi = \frac{S_{ant1} \cdot G^{ze}_{ant2} - S_{ant2} \cdot G^{ze}_{ant1}}{G^{az}_{ant1} \cdot G^{ze}_{ant2} - G^{az}_{ant2} \cdot G^{ze}_{ant1}}. \qquad (1)$$

## Weighting scheme

Due to the measurement with three antennas in spatial coincidence the electric field vector can be reconstructed three times with all possible combinations of measurements from two antennas. Thus in total three values of the same electric field vector are available. This gives rise to the question on the best use of these redundant reconstructions of the electric field vector. Several methods have been developed and tested, for further details consider ref. [12]. In this paper we briefly discuss the so-called gain method, as with this method the best results were achieved. Hereby we do a weighted averaging of the three electric field vectors. As a weight we used the combined simulated sensitivity of the two antennas used to reconstruct the electric field vector. In general the vertical antennas get higher weights the more inclined the air shower was, whereas the horizontally aligned antennas get high weights the more vertical the air shower arrived. With the gain method most events could be reconstructed when including measurements with vertically aligned antennas.

## MEASUREMENTS WITH VERTICALLY ALIGNED ANTENNAS

Including measurements with vertically aligned antennas leads to fewer reconstructible events in radio. One reason for this is the vertical component which is in most cases the weakest (at least at the LOPES site). Another reason could be background noise, as the noise was never studied in the different components separately. The vertical component is expected to be stronger affected by noise in particular with rising zenith angle of the incoming air shower [13]. This behaviour could be confirmed by measurements. By using LOPES as a digital radio interferometer the radio signal can be analysed according to the arrival direction. In figure 2 (lhs) the zenith angle dependence of the background noise in the vertical component is shown. It can be seen that as expected the background noise increases with rising zenith angle. This is harming the efficiency of measurements with the vertical antennas as they get more sensitive and additionally the expected signal gets higher with rising zenith angle. In other words, the best geometries for the measurements with vertically aligned antennas when looking at the sensitivity and predicted emission are the geometries with highest background noise. For the background noise in the north-south and east-west component no significant zenith angle dependence was found.

## SENSITIVITY TO INCLINED SHOWERS

As discussed above, measurements with vertical antennas are supposed to be more efficient the more inclined the air shower was. Thus we compare the reconstruction efficiency with and without vertical measurement as a function of the zenith angle. In figure 2 (rhs) the amount of reconstructible events is shown, in red when including the vertical measurements, in blue when ignoring them. In dark gray the triggers LOPES received from KASCADE-Grande [14, 15] are shown. It can be seen that in general more events can be reconstructed when ignoring the vertical measurements. This is due to the weak signal and the noise in the vertical component. However this figure changes for inclined geometries, as for these more showers can be reconstructed when including vertical measurements. When looking just for events with a zenith angle $\geq 60°$, in total 22 events can be reconstructed when using only horizontally aligned antennas whereas 29 events can be reconstructed when including the vertical measurements LOPES 3D and especially this analysis suffers from the poor statistics caused by a rise in the background noise [12].

## CONCLUSION

In this article we have discussed the benefits of additional measurements with vertically aligned antennas. In a detailed study an increase of background noise with zenith angle in the vertical component was found at the LOPES site. This led to a decrease in the number of reconstructible events when including the measurements with vertically aligned antennas for showers with zenith angles smaller than $60°$. Nevertheless, for showers with zenith angles larger than $60°$ the number of reconstructible events increased when including measurements with vertically aligned antennas. This shows that the measurement with vertically aligned antennas is of importance when it comes to inclined shower detection but the full potential of these measurements can be exploited best when going to radio quiet areas.

## ACKNOWLEDGMENTS

LOPES and KASCADE-Grande have been supported by the German Federal Ministry of Education and Research. KASCADE-Grande is partly supported by the MIUR and INAF of Italy, the Polish Ministry of Science and Higher Education and by the Romanian Authority for Scientific Research UEFISCDI (PNII-IDEI grant 271/2011). This research has been supported by grant number VH-NG-413 of the Helmholtz Association. The present study is supported by the 'Helmholtz Alliance for Astroparticle Physics - HAP' funded by the Initiative and Networking Fund of the Helmholtz Association, Germany an by KSETA, the KIT graduate school for elementary particle and astroparticle physics